\documentclass[10pt,conference]{IEEEtran}
\IEEEoverridecommandlockouts
\usepackage{cite}
\usepackage{amsmath,amssymb,amsfonts}
\usepackage{algorithmic}
\usepackage{graphicx}
\usepackage[]{footmisc}
\usepackage{textcomp}
\usepackage{xcolor,color,colortbl}
\usepackage{tcolorbox}
\usepackage[utf8]{inputenc}
\usepackage{xspace}
\usepackage{listings}
\usepackage{url}
\usepackage{adjustbox}
\usepackage{multirow}
\usepackage{hyperref}
\usepackage{float}
\usepackage{cleveref}
\usepackage[font=small,skip=0pt]{caption}
\usepackage[htt]{hyphenat}
\usepackage{tikz}
\def\BibTeX{{\rm B\kern-.05em{\sc i\kern-.025em b}\kern-.08em
    T\kern-.1667em\lower.7ex\hbox{E}\kern-.125emX}}
\begin{document}

\newcommand{\capgen}{CapGen\xspace}
\newcommand{\major}{MAJOR\xspace}
\newcommand{\pit}{PITest\xspace}
\newcommand{\memu}{MeMu\xspace}
\newcommand{\bddbddb}{\texttt{bddbddb}\xspace}

\newcommand{\defectsj}{Defects4J\xspace}
\newcommand{\corebench}{CoREBench\xspace}

\newcommand{\gv}{G\&V\xspace}

\newcommand{\numBenchmarkProgs}{12\xspace}
\newcommand{\memuAvgSpeedUpPercent}{18.15\%\xspace}
\newcommand{\memuMinSpeedUpPercent}{-0.66\%\xspace}
\newcommand{\memuMaxSpeedUpPercent}{51.77\%\xspace}
\newcommand{\pitSpendsInTopExpensive}{43.21\%\xspace}
\newcommand{\pitSpendsInMutationTesting}{69.97\%\xspace}

\newcommand{\info}[1]{\circled{\textsf{\scriptsize{#1}}}}
\newcommand{\proc}[1]{\fbox{\textsf{\scriptsize{#1}}}}

\crefformat{section}{\S#2#1#3}
\crefformat{subsection}{\S#2#1#3}
\crefformat{subsubsection}{\S#2#1#3}

\definecolor{darkblue}{rgb}{0.0,0.0,0.6}
\lstdefinelanguage{XML}
{
  morestring=[b]",
  morestring=[s]{>}{<},
  morecomment=[s]{<?}{?>},
  stringstyle=\color{black},
  identifierstyle=\color{cyan},
  keywordstyle=\color{darkblue},
  showstringspaces=false,
  basicstyle={\small\ttfamily},
  morekeywords={artifactId,version,groupId,plugin,configuration,failingTests}%
}

\lstdefinelanguage{CSV}
{
  basicstyle={\small\ttfamily},
}

\newcommand{\Comment}[1]{}

\newcommand{\ali}[1]{\textcolor[rgb]{0.0,0.0,1.0}{#1}}
\newcommand{\andi}[1]{\textcolor[rgb]{1.0,0.0,0.0}{AM: #1}}
\newcommand{\an}[1]{\textcolor[rgb]{1.0,0.0,0.0}{AM: #1}}
\newcommand{\re}{\textcolor[rgb]{1.0,0.0,0.0}{[REF]}}

\newcommand{\ie}{\textit{i.e.},\xspace}
\newcommand{\eg}{\textit{e.g.},\xspace}
\newcommand{\etc}{\textit{etc.}\xspace}
\newcommand{\etal}{\textit{et al.}\xspace}
\newcommand{\aka}{a.k.a.\xspace} 
\newcommand*\circled[1]{\tikz[baseline=(char.base)]{
            \node[shape=circle,draw,inner sep=2pt] (char) {#1};}}
            
\definecolor{MistyRose}{rgb}{1.0,0.89,0.88}

\title{Toward Speeding up Mutation Analysis \\by Memoizing Expensive Methods}

\author {
    \IEEEauthorblockN{Ali Ghanbari\qquad\qquad Andrian Marcus}
    \IEEEauthorblockA{
    \textit{University of Texas at Dallas, TX 75080, USA}\\
    \{ali.ghanbari,amarcus\}@utdallas.edu}
}

\maketitle

\begin{abstract}
Mutation analysis has many applications, such as assessing the quality of test cases, fault localization, test input generation, security analysis, etc.
Such applications involve running test suite against a large number of program mutants leading to poor scalability.
Much research has been aimed at speeding up this process, focusing on reducing the number of mutants, the number of executed tests, or the execution time of the mutants.

This paper presents a novel approach, named \memu, for reducing the execution time of the mutants, by memoizing the most expensive methods in the system.
Memoization is an optimization technique that allows bypassing the execution of expensive methods, when repeated inputs are detected.
\memu can be used in conjunction with existing acceleration techniques.
We implemented \memu on top of \pit, a well-known JVM bytecode-level mutation analysis system, and obtained, on average, an \memuAvgSpeedUpPercent speed-up over \pit, in the execution time of the mutants for \numBenchmarkProgs real-world programs. 

These promising results and the fact that \memu could also be used for other applications that involve repeated execution of tests (\eg automatic program repair and regression testing), strongly support future research for improving its efficiency.
\end{abstract}


\section{Introduction}\label{sec:introduction}
Mutation analysis/testing \cite{demillo1978hints} is a program analysis technique that involves generating a pool of program variants, called \textit{mutants}, by systematically mutating (\eg replacing an arithmetic operator with another) program elements and running the test suite against the mutants.
Mutation analysis has been mainly used for assessing test adequacy by computing a \textit{mutation score}, which indicates how good a test suite is for detecting bugs \cite{demillo1978hints,ammann2016introduction,visser2016makes}.
In addition, mutation analysis has been used for other purposes, such as fault localization \cite{wong2016survey,papadakis2015metallaxis,papadakis2012using}, automatic program repair \cite{apr2019cacm,debroy2010using,arcuri2011evolutionary,ghanbari2019practical,le2017s3}, test generation \cite{fraser2015achieving,souza2016strong,demillo1991constraint} and prioritization \cite{shin2019empirical}, program verification \cite{galeotti2015inferring,groce2015verified}, \etc

Despite its success in some practical use cases \cite{coles2016pit,ahmed2017applying}, mutation analysis suffers from poor scalability.
One main reason behind this problem is that the generated mutants must be tested against the test suite, and usually a large number of mutants are generated, making the process lengthy.
Much research has been devoted to reducing the cost of mutation analysis \cite{pizzoleto2019systematic,jia2010analysis},
focusing primarily on: 
(1) reducing the number of generated \cite{untch1993mutation,mateo2012mutant,wong1995reducing} or executed mutants\cite{zhang2016predictive,mao2019extensive,devroey2017automata}; 
(2) reducing the number of tests \cite{chen2018speeding,zhang2012regression,gligoric2010mutmut} or reordering them \cite{zhang2013faster}; 
(3) reducing mutant execution time \cite{king1991fortran,wang2017Faster,tokumoto2016muvm,JustEF2014}.

In this paper, we focus on reducing mutant execution time.
The common aspect of most of the approaches in this category is that they focus, one way or another, on the mutated code.
We contend that the execution time during mutation analysis can also be reduced by reducing the execution time of unmutated code.
Such a speed-up technique will complement existing acceleration techniques, as they are orthogonal to each other.
Specifically, we focus on reducing the execution time of (unmutated) \textit{expensive} methods, \ie those that have a longer execution time relative to other methods.
Given that mutation analysis requires many repeated test executions, and a mutation involves small (usually single-pointed) changes to the program, we expect that unmutated expensive methods are executed frequently.
The more frequently these methods are executed, the bigger the time savings will be.
In support of our idea, an empirical study (see \cref{sec:motivation}) revealed that the execution of the top 20\% most expensive methods account for \pitSpendsInTopExpensive of the  mutant testing execution time (in average).

This paper investigates the use of memoization \cite{michie1968memo} for speeding-up the execution the of expensive methods, in the context of mutation analysis.
Memoization is an optimization technique that stores the results of expensive function calls and returns the cached result when the same inputs occur again, and it has been successfully used for
speeding up recursive functions \cite{cormen2009introduction,michie1968memo}, 
optimizing functional programs \cite{xu2007dynamic,heydon2000caching}, 
and eliminating performance bottlenecks \cite{della2015performance,guo2011using}.
We introduce and evaluate a technique, named \memu (\textbf{Me}moized \textbf{Mu}tation analysis), for reducing the execution time of expensive methods during mutation analysis \textit{via} memoization.

Specifically, after identifying the expensive methods, \memu records a snapshot of the state of the unmutated program at the entry and exit point(s) of the those methods, in the form of input-output pairs and stores them in a \textit{memo-table}. 
When testing the mutants, upon the invocation of an expensive method, \memu does a light-weight table look-up to check if a given input has already been recorded in the memo-table. 
If a match for the given input is found, then it updates the system state with the pre-recorded state, without executing the expensive method.
Otherwise, if the input is not in the memo-table (\ie a \textit{cache miss} occurs), the method is executed.

\memu is independent of  mutant generation or test case selection/reordering and it is meant to be used in conjunction with any existing mutation analysis tool.
We implemented and evaluated an instance of \memu, built on top of the \pit mutation analysis system \cite{delahaye2015selecting}.
As such, the \memu prototype is usable with JVM-based programming languages. 

We used \memu for analyzing the tests of \numBenchmarkProgs real-world programs, resulting
in \memuAvgSpeedUpPercent speed-up over \pit, in average (min. \memuMinSpeedUpPercent, max. \memuMaxSpeedUpPercent), for mutant testing. 

For any mutation analysis optimization technique, speed-up comes with two main challenges: (1) limiting the overhead costs; and (2) maintaining true value of mutation score.
Our work highlights challenges and solutions in achieving these goals.
For example, memoizing all the expensive methods results in significant runtime overhead caused by loading and deserializing large memo-table databases and a large number of cache misses.
We also found that memoizing non-deterministic methods adversely affects the mutation score.
We introduce a novel technique, that we call \textit{provisional memoization} (see \cref{sec:framework}), to reduce the size of the memo-table databases and the number of cache misses.
Provisional memoization also identifies certain non-memoizable methods, \eg those that involve non-determinism.

We argue that the use of memoization (with provisional memoization) for speeding-up mutation analysis is promising and we anticipate that future research will further reduce the overhead.
Such research is worthwhile pursuing, as \memu could also be used for speeding up other automated software quality assurance techniques (\eg \cite{baresi2010testful,gligoric2015regression,apr2019cacm}), which also rely on the repeated execution of a test suite on the program. 

\section{Motivational Empirical Study}\label{sec:motivation}
We conducted an empirical study to understand how much time is used on the repeated execution of the expensive methods, during mutation testing. 
The premise of our memoization approach is that the execution of the most expensive methods amount to a significant percentage of the total mutant execution time.
We used \pit \cite{coles2016pit}, a state-of-the-art JVM bytecode-based mutation analysis system.
It offers 29 mutation operators (including commonly-used ones \cite{ammann2016introduction}) and performs on-the-fly mutation generation and testing \textit{via} ASM \cite{bruneton2002asm} and Java instrumentation API \cite{bib:javaAgent}, mitigating the compilation and test isolation overhead.

As subjects, we selected \numBenchmarkProgs real-world programs (see Table \ref{tab:mainTable}), which are widely used in mutation analysis research \cite{schuler2009efficient,just2011major,zhang2016predictive}. 
Table \ref{tab:mainTable} lists the programs, the revisions that we used, and their sizes (number of tests and methods) .

We measured the time it took to generate and test (execute) the mutants.
To measure mutant execution time, we calculated the difference between time before and after executing the mutant.
By subtracting it from total mutation analysis time, we obtain an approximation of other activities (\eg mutant generation) performed by \pit.
To calculate the execution time of individual methods during mutant execution, we instrumented mutants and injected \textit{before} and \textit{after advises} for using ASM to calculate the difference between time at the entry and exit point(s) of the methods.
We used a Dell Workstation with 3.70 GHz CPU and 126 GB of RAM running Ubuntu 18.04.4 LTS.
All time measurements are in seconds and are the result of the average of two executions rounded to the nearest integer.

Table \ref{tab:mainTable} reports the execution time of the top 20\% most expensive methods, the execution time of all the methods, and the total time needed by \pit to perform the mutant testing and other activities such as, mutant generation, mutation score computation, \etc

We observe that \pit spends, on average, \pitSpendsInMutationTesting of its time on testing mutants.
More importantly, executing the top 20\% most expensive methods in the programs accounted, in average, for \pitSpendsInTopExpensive of mutant test execution time (min. 10.11\%, max. 78.06\%).
The findings imply that reducing the execution time of a relatively small fraction of methods (\ie the 20\% most expensive ones) may lead to a significant reduction in the overall mutation analysis time.
They serve as motivation for our approach to focus on method-level optimization for reducing the execution time.

\section{Memoized Mutation Analysis Framework}\label{sec:framework}
\begin{figure}
    \centering
    \includegraphics[scale=0.5]{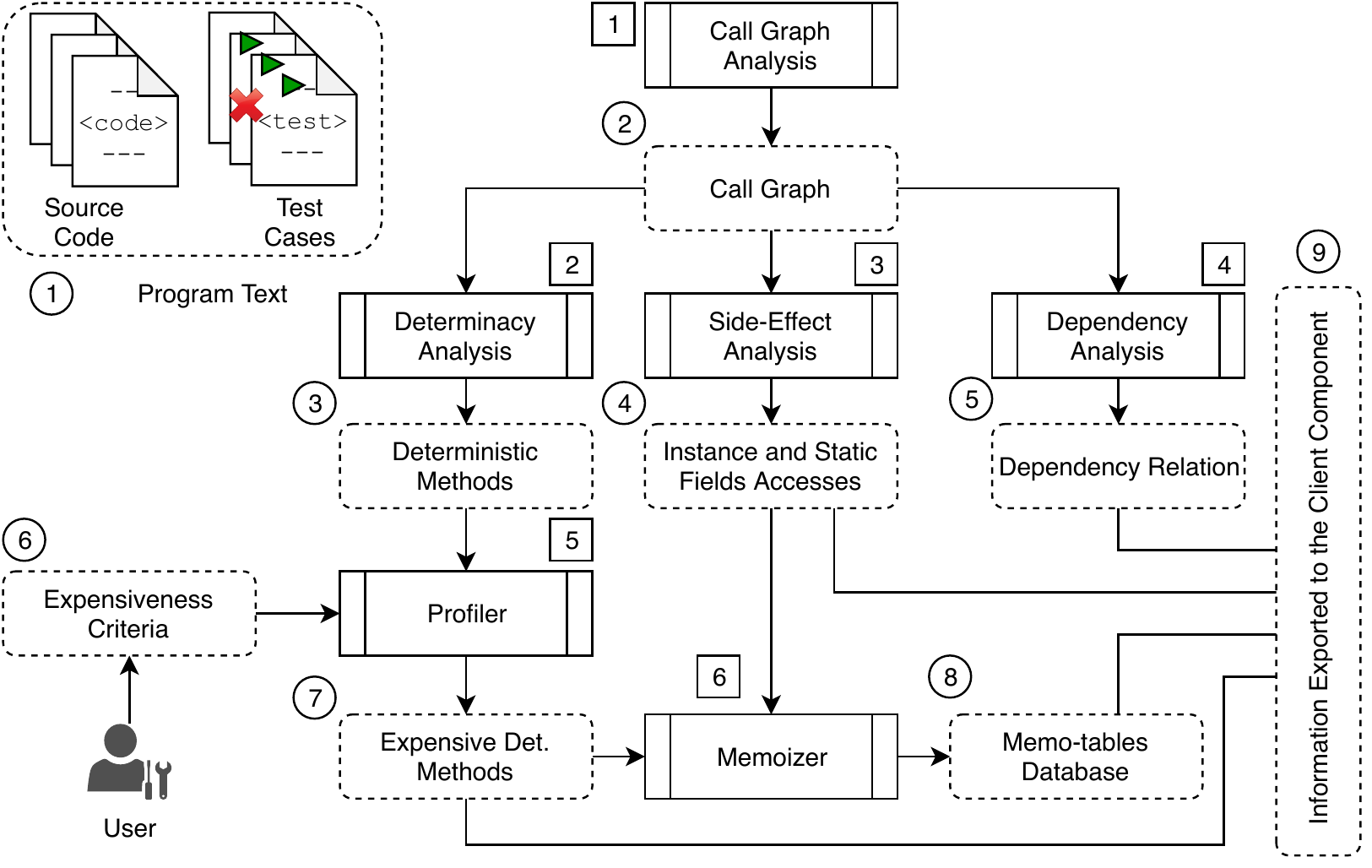}
    \caption{The memoization component of \memu. Processes are represented as double-lined rectangles and information produced/consumed by processes using dashed rounded rectangles.
    Each process uses the program source code and tests as input.}
    \label{fig:memoArch}
\end{figure}

\memu is designed as a framework with two main components: the \textit{memoization} component and the \textit{client} component.
The \textit{memoization component} (see Fig. \ref{fig:memoArch}) is responsible for identifying and memoizing the expensive methods, and passing this information to the client component.
The \textit{client component} can be an existing mutation analysis tool that is modified to intercept the execution of expensive methods  identified by the memoization component so as to check whether or not it can reuse the already computed results instead of re-executing the method.
We implemented the client component for mutation testing by modifying \pit \cite{delahaye2015selecting}.

We describe the data used and produced (denoted by \info{x}) by the processes (denoted by \proc{y}) in the memoization component (see Fig. \ref{fig:memoArch}) and the client component.
In short, to memoize a method, \memu records a snapshot of the state of the unmutated program at the entry and exit point(s) of the the expensive methods, in the form of input-output pairs and stores them in a \textit{memo-table}.
The collection of memo-tables, \ie the \textit{memo-tables database} \info{8}, is then passed to the client component, which uses it to bypass the execution methods, when a ``cache hit'' occurs during mutant execution.

It is impractical to memoize all methods, as it leads to large overhead and in many cases the execution time of a method may be actually shorter than a look-up in the memo-table.
Hence, as we discussed before, we focus on memoizing only the expensive methods.
Given a \textit{program text} \info{1} comprised of the source code and a test suite, the framework needs first to determine which methods to memoize.

\subsection{Which methods to memoize?}
In order to avoid memoizing non-expensive methods, \memu uses two user-provided parameters: a \textit{threshold},  $\tau$, and a \textit{limit}, $\ell$, which define the \textit{expensiveness criterion} \info{6}. 
It attempts to memoize the $\ell$ most expensive methods, with execution time longer than $\tau$ milliseconds.
However, not all of these expensive methods are memoizable. 

The framework applies a \textit{call graph analysis} \proc{1} to obtain the \textit{call graph} \info{2}.
In our implementation, we used the WALA program analysis infrastructure \cite{dolby2015tj} to construct a 0-CFA call-graph, but, of course, other tools may also be used.
The call graph is then used by additional analyses to determine which of the expensive methods should be memoized. 

First, the \textit{dependency analysis} \proc{4} determines the reflexive, transitive closure of the call graph, which is also sent to the client.
The resulting \textit{dependency relations} \info{5} are used for identifying methods that should not be memoized.
If the intercepted method (\ie the one to be memoized) depends on a mutated/modified method or itself undergoes a mutation/modification, then the method shall not be memoized.

Second, the \textit{determinacy analysis} \proc{2} identifies the methods that depend on time and/or random generator or return values computed in such a manner.
We refer to these methods as likely non-deterministic methods.
\memu does not memoize these methods as they might result in large number cache misses (due to the way their input/output is obtained) or change the semantic of programs.
The set of methods that are not likely non-deterministic (\ie \textit{deterministic methods} \info{3} are used in the next process.

Finally, the \textit{profiler} \proc{5} instruments the system to measure the execution time of the likely deterministic methods and determines the \textit{expensive methods} \info{7} that will be memoized.
It also records coverage information of each test case used for excluding unnecessary test cases, for faster memoization.

\subsection{Memoization}
Recording all variables in a memo-table may lead to very large tables.
So, before constructing the memo-tables, \memu filters out fields that are untouched.
The \textit{side-effect analysis} \proc{3} determines which method may access which static/instance fields, either directly or by calling another method.
To keep the size of memo-tables small and optimize table look-up within the client, the framework only uses the \textit{accessed fields} \info{4}.
The \textit{Memoizer} \proc{6} constructs a minimal \textit{memo-tables database} \info{8} of the methods that are deemed memoizable in the previous steps (\ie the expensive, deterministic methods).
This is done by applying two filtering steps.

First, \memu determines which methods will not result in failures when memoized. 
This is achieved via provisional memoization which tentatively memoizes methods and excludes non-memoizable ones.
Specifically, we consider a memoization attempt on a method as failed if memoizing the method results in (new) failed tests.
In this way, we can single out non-memoizable methods.
Second, before passing the \textit{memo-tables database} to the client component, \memu removes the methods incurring cache misses when they are tested against covering tests.
This is done by post-processing the database using the execution information obtained during provisional memoization.

\subsection{Client Component}
The client component for \memu is constructed by modifying \pit such that it loads the memo-tables database in each mutant testing process that \pit forks, and we instrument the mutant code such that the memoizable methods do a light-weight check before proceeding running their bodies.
The methods check if they are mutated or depend on some mutated method.
If that is the case, no memoization shall take place.
Otherwise, they do a light-weight table look-up based on the state of the system at their entry points and update the system state if such a state have occurred previously.
Then, the method immediately returns without executing its body.


\begin{table*}[ht!]
    \centering
    \caption{Result of applying \pit and \memu on \numBenchmarkProgs systems. \textbf{MT}=mutant testing time, \textbf{Score}=mutation score,  \textbf{\#MM}=memoized methods.}
    \label{tab:mainTable}
    \begin{adjustbox}{width=0.85\textwidth}
        \begin{tabular}{|c|c|r|r|r|r|r|c|r|r|r|r|}
            \hline
            \multicolumn{4}{|c|}{\textbf{Subject Information}} & \multicolumn{4}{c|}{\textbf{\pit Execution Information}} & \multicolumn{4}{c|}{\textbf{\memu Execution Information}} \\ \hline
            \textbf{Project Name} & \textbf{Rev} & \textbf{\#Test} & \textbf{\#Method} & \textbf{Total (s)} & \textbf{MT All (s)} & \textbf{MT Top \%20 (s)} & \textbf{Score} & \textbf{\#MM} & \textbf{MT (s)} & \textbf{\#Hit} & \textbf{\#Miss} \\ \hline \hline
            \textsf{commons-codec} & \texttt{5ef5} & 851 & 792 & 730 & 418 & 283 & 0.863341 & 21 & 220 & 933,291 & 45 \\ \hline
            \textsf{commons-math} & \texttt{0da6} & 5,246 & 8,364 & 47,460 & 37,429 & 17,352 & 0.781098 & 41 & 36,280 & 81,788 & 8,638 \\ \hline
            \textsf{commons-cli} & \texttt{0b45} & 390 & 292 & 140 & 55 & 20 & 0.897196 & 5 & 47 & 8,962 & 431 \\ \hline
            \textsf{commons-csv} & \texttt{f368} & 306 & 189 & 402 & 282 & 114 & 0.841484 & 3 & 261 & 3,998 & 0 \\ \hline
            \textsf{closure-compiler} & \texttt{1dfa} & 7,907 & 10,536 & 64,860 & 62,665 & 30,157 & 0.779994 & 5 & 59,493 & 2,246 & 54 \\ \hline
            \textsf{commons-io} & \texttt{2ae0} & 132 & 1,325 & 2,329 & 1,682 & 170 & 0.805581 & 1 & 1,641 & 2,406 & 187 \\ \hline
            \textsf{commons-fileupload} & \texttt{047f} & 82 & 306 & 275 & 232 & 160 & 0.606796 & 4 & 170 & 3,247 & 29 \\ \hline
            \textsf{jfreechart} & \texttt{2266} & 2,201 & 9,110 & 3,156 & 1,565 & 164 & 0.346555 & 0 & N/A & N/A & N/A \\ \hline
            \textsf{commons-imaging} & \texttt{fd01} & 93 & 2,439 & 4,740 & 3,954 & 785 & 0.422943 & 7 & 3,407 & 475 & 46 \\ \hline
            \textsf{commons-lang} & \texttt{687b} & 2,295 & 2,596 & 1,926 & 761 & 594 & 0.864125 & 4 & 367 & 119 & 12 \\ \hline
            \textsf{joda-time} & \texttt{9a62} & 4,043 & 4,366 & 1,839 & 1,362 & 680 & 0.468076 & 7 & 1,371 & 0 & 0 \\ \hline
            \textsf{commons-geometry-euclidean} & \texttt{b36d} & 1,628 & 1,934 & 6,780 & 6,390 & 2,696 & 0.942488 & 50 & 4,595 & 85,709 & 3,350 \\ \hline
        \end{tabular}
    \end{adjustbox}
\end{table*}


\section{Empirical Evaluation and Discussion}\label{sec:results}
We conducted an empirical study to assess whether \memu obtains any speed-up in mutant execution time compared to \pit.
We used the same subjects as in the motivational study, described in \cref{sec:motivation}.
We set $\tau$ and $\ell$ to 1 ms and 20\% of number of methods for each subject program, respectively.

The right hand side of Table \ref{tab:mainTable} summarizes the information about \memu's execution.

Comparing \pit's and \memu's mutant testing time (\ie the two MT (s) columns in Table \ref{tab:mainTable}),
in 10 out of 12 cases, \memu completes the execution faster.
Excluding \textsf{\small{jfreechart}}, \memu results in \memuAvgSpeedUpPercent speed-up (on average - minimum \memuMinSpeedUpPercent, maximum \memuMaxSpeedUpPercent) over \pit.

We analyzed the two cases where \memu did not obtain a speed-up: \textsf{\small{jfreechart}} and \textsf{\small{joda-time}}. For the \textsf{\small{jfreechart}} system, our implementation of \memu fails to completely restore the system state, so provisional memoization fails to memoize any methods, so we did not perform memoized mutation testing for that subject, hence the "N/A" values the table.
The reason is that \textsf{\small{jfreechart}} uses graphic libraries that involve system states, which are inaccessible through the Java reflection API \cite{reflectionAPI} used by our framework.
We expect that \memu has the same problem with other similar systems.
However, this is not a shortcoming of the idea, rather a consequence of our engineering choice to use reflection and will be addressed in future work.

Thanks to the provisional memoization algorithm, we have been able to exclude non-memoizable methods (see the \#MM column).
On average, the 11 system (not counting \textsf{\small{jfreechart}}) have 3,521 methods (20\% of which is 704).
\memu memoizes, no more than 1\% of the methods (min. 1, max. 50), yet it results in a considerable amount of time saving.

Provisional memoization ensured that the number of cache misses (\#Miss column) is smaller than that of cache hits (\#Hit column) for the memoized methods.
However, the algorithm is not perfect; for the \textsf{\small{joda-time}} system, \memu is slower than \pit, because the memoized methods do not result in any cache hits during the mutant executions.

Finally, we believe the memoization also resulted in a lossless mutation testing, because for the subject programs with constant mutation scores between runs, the mutation score before and after memoization did not change.

\section{Related Work}\label{sec:related}
Conventionally, we classify approaches for reducing mutation analysis costs into three major categories \cite{offutt2001mutation,pizzoleto2019systematic}: (1) \textit{do fewer} approaches strive generating/testing as few mutants as possible with minimal adverse effect on mutation score \cite{wong1995reducing,zhang2016predictive,mao2019extensive,chen2018speeding,devroey2017automata}; (2) \textit{do faster} approaches are meant to generate and run mutants as fast as possible (without any concern about mutation score) \cite{untch1993mutation,mateo2012mutant,zhang2013faster,zhang2012regression,gligoric2010mutmut,howden1982weak,durelli2012toward}; (3) \textit{do smarter} approaches are intended to distribute the workload of testing mutants into several machines or several cores of a single machine \cite{gopinath2016topsy,li2015mutation,coles2016pit}, or factor out shared state between mutant executions and avoid re-executing them \cite{king1991fortran,wang2017Faster,tokumoto2016muvm,JustEF2014}. 

\memu fits in the third category as it applies a semantic-preserving program optimization method (\ie memoization in this case) on the unmutated parts of the mutants to avoid re-executing (expensive) methods for which the state of the system at the entry and exit point(s) do not change from one execution to another.
We discuss here the works in this category, which we consider most related to \memu.

Split-stream \cite{king1991fortran,tokumoto2016muvm} and its modern variants are intended to avoid repeated execution of part of the code that is shared between mutants.
Mutations targeting the same program element, result in many mutants that share the same code before the mutation impact point.
Executing this portion of the mutants (provided that the program is deterministic) will always result the same output.
Split-stream  runs these portions only once and fork different processes for the each mutant \emph{after} the mutation point of impact to test individual mutants.
The modern incarnation of split-stream \cite{wang2017Faster} attempts to reuse shared program states even after mutation point of impact.

Just \etal \cite{JustEF2014} propose three runtime optimizations that result in 40\% speed up of their \major mutation analysis system \cite{just2011major}: (1) if a mutation does not result in program state change immediately after the mutation point, it marks the corresponding mutant as \emph{survived}, \ie not killed, and terminates the test execution; (2) even if a mutation infects the system state in an expression while the change does not propagate to the subsequent statements, it marks the corresponding mutant as survived and terminates the test execution; (3) mutants that infect the state of the system in the same way should only be executed once.

Since \memu optimizes the execution of unmutated code and the memoization does not influence the effect of the mutation, it can complement existing cost reduction techniques.
The information collection processes can be parallelized with the pre-processing done by such complementary techniques, in a non-interfering manner, to further speed-up the mutation analysis process.

\section{Conclusions and Future Work}\label{sec:conclusions}
The new idea put forward in this paper is speeding up mutation analysis by automatically memoizing expensive methods.
Our optimization is orthogonal to the state-of-the-art cost reduction techniques for mutation analysis and can be used together with them to further speed up the process.
An empirical study using state-of-the-art, JVM-based, mutation analysis tool \pit \cite{coles2016pit} and \numBenchmarkProgs real-world  programs, revealed that
\pitSpendsInTopExpensive (avg.) of the mutant execution time is spent on executing the top 20\% most expensive methods.
This finding supports the intuition behind the memoization-based approach for speeding up mutant execution.
An additional empirical study showed that memoizing a small  subset of these expensive methods (1\% of all methods) leads  to an average of \memuAvgSpeedUpPercent speed-up during mutant testing.
We uncovered two specific issues, important for the successful memoization: identifying non-memoizable methods and minimizing the number of cache misses during testing.
Provisional memoization shows promise in tackling these issues.
Future work will focus on more light-weight techniques, based on statistical models, which may be less costly than provisional memoization.
The other analyses used during memoization can be optimized through parallelization.

We contend that other software quality assurance methods that rely on repeated execution of the code,  such as, automatic program repair and regression testing, can also benefit from the memoization idea.
Hence, the potential advantages largely exceed those reported here, supporting future work that will further optimize the memoization approach.

\section*{Data Availability}
Data are available at \url{https://bit.ly/3omErsz}.
\section*{Acknowledgments}
This research was partially supported by the NSF grants CCF-1910976 and CCF-1955837.

\bibliographystyle{IEEEtran}
\bibliography{bibdb}
\end{document}